\documentclass[letter,traditabstract]{aa}

\usepackage{graphicx}
\usepackage{txfonts}
\usepackage{natbib}
\bibpunct{(}{)}{;}{a}{}{,} 

\begin{document}


\title{Highly extinguished host galaxy of the dark GRB 020819}

\author{A. K\"upc\"u Yolda\c{s}\inst{1} 
\and J. Greiner\inst{2}
\and S. Klose\inst{3}
\and T. Kr\"uhler\inst{2,4}
\and S. Savaglio\inst{2}}

\institute{European Southern Observatory, Karl-Schwarzschild-Str. 2, 85748 Garching, Germany\\ \email{ayoldas@eso.org}
\and Max-Planck-Institut f\"{u}r extraterrestrische Physik, Giessenbachstrasse 1, 85748 Garching, Germany
\and Th\"uringer Landessternwarte Tautenburg, Sternwarte 5, D-07778 Tautenburg, Germany
\and Universe Cluster, Technische Universit\"{a}t M\"{u}nchen, Boltzmannstrasse 2, 85748 Garching, Germany}


\date{Received 27 April 2010 / Accepted 06 May 2010}

\abstract{
We analyse the properties of the host galaxy of the optically dark gamma-ray burst (GRB) 020819 (z = 0.41) and discuss the possible implications in the context of ``dark" GRBs.
We present {\it g$^\prime$r$^\prime$i$^\prime$z$^\prime$JHK} photometry of the host galaxy and fit the broad spectral energy distribution including the public Spitzer IRAC data using stellar population models.
The broad spectral energy distribution (SED) indicates a high extinction, A$_V \sim$1.8 -- 2.6 mag,  for this relatively massive galaxy.
This is the highest global extinction for a GRB host galaxy with a robust spectroscopic redshift. The properties of the host galaxy  
are indicative of dusty, intense star-formation, which differ from those of the current sample of GRB hosts. This implies that the dust extinction is one of the main reasons for the darkness of low-redshift bursts and that the long GRB host population is far more diverse than previously anticipated.}


\maketitle 

\section{Introduction}

The afterglows of gamma-ray bursts (GRBs) were predicted by the canonical fireball theory \citep[e.g.,][]{mes97} and have since been detected for more than a decade \citep{vanpar97}. The optical-to-X-ray flux ratio of a GRB afterglow is determined by the fireball theory \citep[e.g.,][]{spn98}, where the optical-to-X-ray spectral index of a canonical GRB afterglow is expected to be $\beta_{OX} \ge $ 0.5 \citep{jak04}. However, for a significant proportion (25\% - 50\%) of all well-localised GRBs, no optical/near-infrared afterglow is detected and/or the optical afterglow emission is lower than that expected from the X-ray afterglow emission ($\beta_{OX} <$ 0.5) \citep[e.g.,][]{fyn09,cen09,dep06}. These bursts are called ``dark" GRBs. The nature of the dark bursts is still to be understood, although several ideas have been proposed to explain why some bursts are dark in the optical bands: (i) shifting of the rest-frame optical afterglow emission and the Lyman-limit towards longer wavelengths for high redshift bursts, (ii) intrinsic dimness of the afterglow (for those with no optical emission and unknown $\beta_{OX}$), or (iii) high extinction in either the host galaxy or the circumburst environment and along the line of sight \citep[e.g.,][]{fyn01,laz02,dep03,fox03}. 

GRB 020819\footnote{This burst is in fact GRB 020819B (see Jochen Greiner's web site http://www.mpe.mpg.de/$\sim$jcg/grb020819.html). However it is referred to as GRB 020819 in all related GCNs and published papers, and we therefore also refer to it as GRB 020819 to avoid further confusion.} was detected by the High Energy Transient Explorer (HETE) satellite and identified as a moderately bright long-soft burst with T$_{90}\sim$20 seconds and a peak brightness of $\sim$5 crab \citep{gcn1508}. Follow-up ground-based observations led to a $K^\prime$-band limit of 19.5 mag and R-band limit of 22.15 for the afterglow emission $\sim$9 hrs after the burst \citep{klose03,gcn1517}. Therefore, GRB 020819 was classified as a dark burst. 

\begin{figure}[h]
\centering
\includegraphics[scale=0.6]{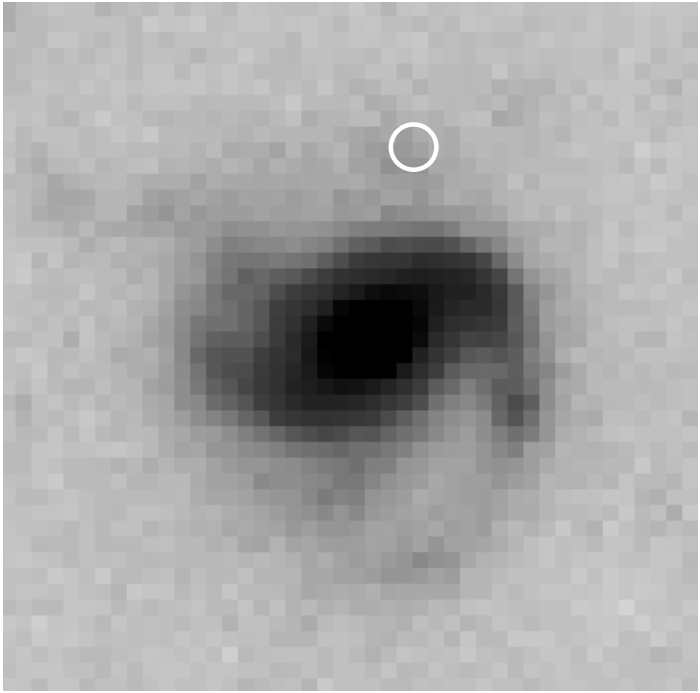}
\caption{VLT/FORS2 $R$-band image of the host galaxy of GRB 020819 taken on 15 Sep 2002. It is clearly resolved as a barred spiral galaxy with bright knots of star-formation in its spiral arms and with its shape possibly resembling NGC 1300. The approximate position of the radio transient is shown by the circle (following Jakobsson et al. 2005). The field of view is $11\,\times\,11$ arcsec$^2$. North is up, east is left.}
\label{fig:fors}
\end{figure}

\cite{gcn1842} later discovered a declining radio source consistent with radio afterglow emission located at a faint blob 3\arcsec   away from the centre of a spiral galaxy \citep{jak05}. This galaxy, at z=0.41, is identified as the host galaxy of GRB 020819 with a chance coincidence probability of 0.8\% \citep{jak05}. The host has a visible extent of $\sim$6\arcsec  -- 7\arcsec  (32.5 -- 38 kpc) in the $R$-band (see Fig.~\ref{fig:fors}). 

Here we present broad-band photometric observations of the host galaxy of GRB 020819 (Sec. 2) and the analysis of its spectral energy distribution (Sect. 3). In Sect. 4, we discuss the results and their implications for the dark GRB framework.

\section{Observations}

The host galaxy of GRB 020819 was observed with GROND \citep{jcg08} in {\it g$^\prime$r$^\prime$i$^\prime$z$^\prime$JHK} on 29 October 2007 for $\sim$1 hour at an average airmass of 1.23. The data reduction and aperture photometry of the host was performed using the data reduction and photometry tools of the GROND pipeline \citep[see e.g.][]{aky08} based on IRAF/PyRAF.  Astrometry and photometric calibration was performed using the SDSS catalogue \citep{aba09} for the optical bands, and the 2MASS catalogue \citep{skru06} for infrared bands.  

\begin{table*}[th]
\centering
\caption{The photometry of the host galaxy of GRB 020819 $^a$}
\label{tab:phot}
\begin{tabular}{ccccccc}
\hline\hline
g$^\prime$ & r$^\prime$ & i$^\prime$ & z$^\prime$ & J & H & K \\
\hline
20.31$\pm{0.02}$ & 19.43$\pm{0.06}$ & 19.10$\pm{0.07}$ & 18.73$\pm{0.01}$ & 18.46$\pm{0.04}$ & 18.23$\pm{0.04}$ & 18.46$\pm{0.11}$ \\
\hline
\end{tabular}
\begin{list}{}{}
\item [$^a$]  All magnitudes are in units of AB mag and are corrected for foreground Galactic extinction.
\end{list}
\end{table*}

The brightness of the host galaxy was derived using aperture photometry with a radius of $\sim$4.5\arcsec  (3$\times$PSF) in all bands. The results of the photometry are shown in Table \ref{tab:phot}. All magnitudes are corrected for a foreground Galactic extinction of E(B--V) = 0.069 \citep{schl98}.
We note that our $g^\prime$-band magnitude, 20.31$\pm{0.02}$, is brighter than the $B$-band magnitude, 21.9$\pm{0.5}$, of the host reported by \citet{jak05}. The difference is larger than that expected because of the differences between the two filter curves. The reason for this difference is probably the inaccurate calibration of the Keck $B$-band data where \citet{jak05} used a single faint star for calibration, which was the only available star in the field (Jakobsson, private communication). 

\section{Spectral energy distribution}

The spectral energy distribution (SED) was analysed using HyperZ \citep{bol00}. The redshift of the host was fixed to the spectroscopic value of z = 0.41. We allowed the extinction value A$_V$ to vary and fit the host extinction by using 4 different reddening laws: MW, LMC, SMC, and Calzetti \citep{cal00}. 
To fit the stellar populations, we used two different stellar population model libraries:  the GISSEL library of \citet{bc93} and the stellar population models of \citet{maras05}. The initial mass function of all models follows the Salpeter law \citep{sal55}. 

The GISSEL library contains models with 8 different star-formation histories: an instantaneous burst, a constant star-forming system, and six exponentially decaying star-formation rate (SFR) models where SFR $\propto$ exp(--t/$\tau$), $\tau$ being the characteristic lifetime. There are two sets of stellar population models in the GISSEL library: i) those with metallicities fixed to the solar value, and ii) those for which the metallicity evolve, where the evolution in metallicity of the stellar population is explicitly taken into account so that there is a natural link between age and mean metallicity \citep{bol00}. The ages of the templates vary between zero and the age of the Universe (at z = 0.41 in the case of this GRB) covered with unevenly distributed steps \citep{bol00}.

From \citet{maras05} models, we used both single and composite stellar population models. We selected the models with metallicity Z = 0.04 (twice the solar value) to maintain agreement with the metallicity measured based on the spectroscopic line ratio of [NII]$\lambda$6584/[OII]$\lambda$3727, leading to log(O/H) + 12 = 9.0$\pm{0.1}$ \citep{leve10}.  The single stellar population (SSP) models that we used have a red horizontal branch morphology, meaning that the entire horizontal branch lifetime is spent on the red side of RR Lyrae strip \citep{maras05}.

The host galaxy of GRB 020819 was also observed using IRAC on-board the {\it Spitzer} satellite. \citet{sven10} measured the IRAC 3.6 $\mu$m and 5.8 $\mu$m brightnesses as 18.96$\pm{0.02}$ mag (corresponding to 95.1$\pm{1.8}~ \mu$Jy) and 19.27$\pm{0.22}$ mag, respectively. \citet{ceron08} also reports the IRAC 3.6 $\mu$m band brightness as 97$\pm{2}~ \mu$Jy, in agreement with \citet{sven10}. We used these measurements with our GROND data to form a broad SED that covers the rest-frame UV to infrared wavelengths. In the following paragraphs, we present the SED fits of the host galaxy of GRB 020819, using GROND data combined with the IRAC brightnesses. The errors quoted in the following sections are at the 1 $\sigma$ level and calculated by taking into account all the galaxy templates of a given stellar population library used in that fit.

\subsection{Fits to the GROND and IRAC data}

\begin{figure*}
\centering
\includegraphics[angle=270,scale=0.6]{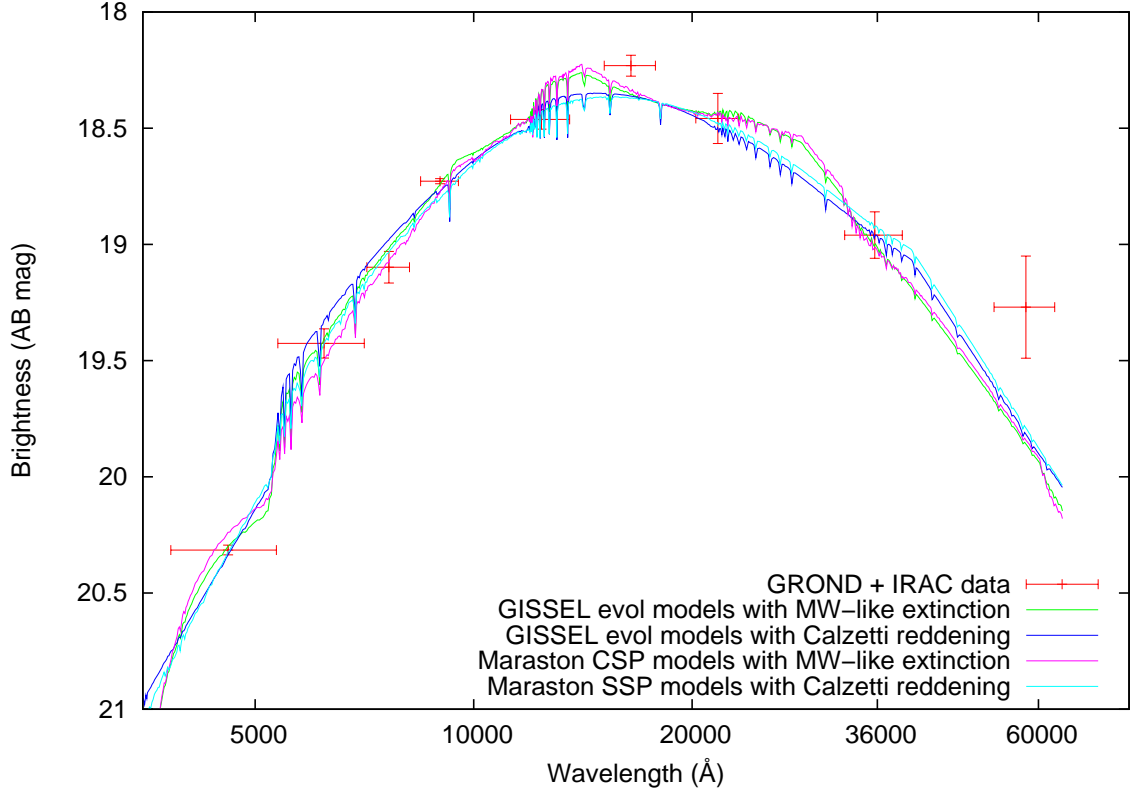}
\caption{The fits to the SED of the host of GRB 020819 using GROND and IRAC data. Note that we repeated the fits after increasing the  IRAC 3.6$\mu$m band brightness error from 0.02 mag to 0.1 mag. Error bars for both cases are shown in the figure.}
\label{fig:wirac_sb_mw}
\end{figure*}

The best-fit models to the broad-range SED formed by GROND and IRAC data, are those with a MW-like extinction for both stellar library models (see Fig.~\ref{fig:wirac_sb_mw}). Using the models with evolving metallicity from the GISSEL library, the best fit galaxy model has an age 30$^{+60.5}_{-15}$ Myr and an extinction of A$_V$ = 2.21$\pm{0.39}$ mag using a MW reddening law and an exponentially decaying SFR with $\tau$=1 Gyr ($\chi^2_{\nu}$ = 2.73). The stellar mass of the best-fit model is $\sim$10$^{10.4}$ solar masses, in agreement with \citet{sandra}.

Using Maraston SSP and CSP models with a fixed metallicity of Z = 0.04, the best-fit is obtained with a CSP model that has an exponentially decaying SFR with $\tau$=1 Gyr, age 6.0$\pm{0.5}$ Myr and A$_V$=2.47$^{+0.26}_{-0.13}$ using a MW reddening law. The models with longer SFR e-folding times, $\tau$, provided similarly good fits with the same age and extinction values as best-fit. All other reddening laws provided a significantly ($>$1 $\sigma$) poorer fit.
SSP models also fit similarly well but only when using a Calzetti reddening law with a best-fit model of A$_V$=3.12 and age = 3 Myr. 

We note that when using a MW reddening law, even the best fit does not provide a good fit to the H-band (see Fig.~\ref{fig:wirac_sb_mw}). Similarly, the stellar population models extinguished with the Calzetti reddening law overestimate the $K$-band flux
. The reason for the imperfect fits to the $H$-band flux is possibly emission lines such as Pa$\beta$ and He I from the ionised interstellar gas, which fall into the $H$-band at the redshift of this galaxy.

\section{Results and discussion}

The SED fits show that the optical and near-infrared emission from the galaxy is dominated by a very young stellar population. The young age of the dominant stellar population prevents us from distinguishing between models with different SFR e-folding times and hence does not allow us to determine the SFR history of the galaxy. The age of the dominant stellar population is $\sim$6 Myr, if the metallicity is fixed to Z = 0.04, and slightly older, $\sim$30 Myr, if the metallicity evolves in proportion to the age of the stellar population.
The MW-like extinction provides better fits to the data than the Calzetti reddening. The best-fit extinction value is in the range A$_V \sim$ 1.8 -- 2.6 mag depending on the galaxy template library used. 

In principle, there is a degeneracy between the dust extinction in a galaxy and the age and metallicity of the mean stellar population of a galaxy, since all three redden the colour of the galaxy. However, it is also known that metal-poor galaxies have less dust content and lower global infrared emission \citep{cal10,can05,can06,wal07}. 
This is supported by our finding that the dust extinction and the age values obtained by fitting the solar/super-solar metallicity models and the evolving metallicity models roughly agree with each other, indicating that either the metallicity is high or it does not have a significant effect on the determination of the dust extinction for this galaxy, or both. In other words, the galaxy is highly extinguished according to the fits by synthetic stellar population models, independent of the metallicity of the models. 

The age and dust extinction values we obtained agree with the spectroscopic measurements of \citet{leve10} who found an age of 7.8$\pm{0.9}$ Myr using the H$_{\beta}$ emission line equivalent width and E (B -- V) = 0.64 using the H$_{\alpha}$/H$_{\beta}$ line ratio. For a MW-like total-to-selective extinction ratio R$_V$ = 3.1, E (B -- V) = 0.64 corresponds to A$_V$ = 1.98 mag, which is in agreement with the A$_V$ values from our SED fits. \citet{leve10} also measured the metallicity of the host galaxy based on the spectroscopic line ratio of [NII]$\lambda$6584/[OII]$\lambda$3727 and found a super-solar metallicity of log(O/H) + 12 = 9.0$\pm{0.1}$. 

The age (and extinction) value obtained from either our SED fits or the spectrum, by \citet{leve10}, do not agree with that of \cite{sandra}. The reason for this disagreement is probably that they fit the SED of this host using only the $B$-, $R$-, and $K$-band magnitudes available in the literature, where the $B$-band magnitude has a large uncertainty (see Sect.~2). Furthermore, their fits are restricted to the A$_V$ range of 0.0 -- 2.0 mag and metallicities no more than the solar value, which do not agree with the best-fit model A$_V$ or metallicity of this host. Hence, our analysis illustrates the importance of accurate photometry, a broad SED coverage, and correctly selected parameter ranges.
 
On the other hand, our best-fit stellar mass of $\sim$10$^{10.4} M_\odot$  agrees with $\sim$10$^{10.50\pm{0.14}} M_\odot$ of \cite{sandra}, indicating that the stellar mass calculations based on the optical/near-infrared SED are much less affected by the choice of parameter ranges. A reason for the robustness of the mass estimation is that the SED covers the Balmer break and the rest-frame near-infrared emission of the galaxy.

The calculation of the dust extinction based on the afterglow emission limits of GRB 020819 in the $R$- and $K^\prime$-band by \cite{klose03} show that if the afterglow is similar to that of GRB 011121 and at the same redshift (z=0.36), then an extinction of A$_V \sim$ 4 mag is required to dim the $R$-band afterglow and A$_V \sim$ 20 mag to dim the $K^\prime$-band afterglow below the limits. 
We performed a fit to the afterglow upper limits of the $R$- and $K^\prime$-band at z = 0.41 using a host galaxy extinction of A$_V$ = 2.2 mag (following a MW-like extinction law) and found that the afterglow should have a spectral decay index $\beta >$0.82. This value is very close to the canonical value of 0.8, showing that the afterglow of this dark GRB is probably quite normal i.e., not significantly different from other long GRB afterglows. Therefore, the ``darkness" of this GRB is probably not related to the properties of its afterglow but the dust in the host galaxy.
The high A$_V$ value is consistent with the ÒdarkÓness of the GRB afterglow, although the A$_V$ that we deduced here is not only for the GRB region but the whole galaxy. 

There are several bursts with high extinction in the circumburst region or along the line of sight, derived using the afterglow emission \citep[e.g.][]{djor01,rol07,jau08,tan08,kann06,kann09,has10}. However, GRB 020819 is the first case with known redshift and a well-constrained high extinction value for the entire galaxy. A similar case is that of GRB 090417B (z=0.345) for which \cite{hol10} measure A$_V\sim$3.5 mag based on the UV and optical SED of this dark GRB host.
The high host-galaxy extinctions of these low-redshift dark GRBs also show that while the dust extinction is one of the main reasons for the darkness of low-redshift bursts, at high-redshift (z$>$1) the dust extinction required to dim the afterglow is not that high.  

The case of GRB 020819 is an exception in the context of both the host galaxies of long GRBs with bright optical afterglows, and the dark GRB hosts. Host galaxy studies of dark GRBs indicate that the majority of the dark GRB hosts are similar to normal long GRB hosts, which do not have very high extinction or very high redshifts \citep{perley09}. However, the host extinction of GRB 020819 is much higher than the average extinction, A$_V \le$  0.5, of a sample of hosts galaxies of long GRBs with detected optical afterglows, obtained either using spectroscopy or SED fitting \citep[e.g.,][]{han10,sandra}. Furthermore, the spectroscopy of the host indicates very high star-formation-rate (SFR) of 23.6 M$_\odot$/yr \citep{leve10} corresponding to a specific SFR (SFR / stellar mass) of 0.94 per Gyr$^{-1}$ using our best-fit stellar mass 10$^{10.4} M_\odot$. This specific SFR is higher than the median specific SFR $\sim$0.8 Gyr$^{-1}$ of long GRB host galaxies \citep{sandra}. 
These properties mark the host galaxy of GRB 020819 as a galaxy with dusty, intense star-formation, distinguished from the currently known GRB host population.

The host galaxy of another dark GRB 051022 (z=0.807) shows similarly exceptional properties in terms of mass ($\sim$ 10$^{10.4} M_\odot$), SFR ($\sim$ 36 M$_\odot$/yr) and metallicity (log(O/H) + 12 = 8.77) \citep{sandra,gra09}. We note however that there is a large uncertainty in the metallicity calculation for this host depending on whether the upper branch solution calibrated with metal-rich galaxies or a lower branch solution calibrated with metal-poor galaxies is used \citep[see e.g.,][]{kob99,kew02}. 
The extinction of the host galaxy of GRB 051022 \citep[A$_V$=1.55, ][]{leve10} is also relatively high compared to the average long-GRB host population. 
Similarly, there are a few dark GRBs that have both high extinction and near-solar metallicities derived using afterglow observations \citep[e.g.,][]{wat06,kru08,eli09,pro09}, indicating that hosts such as that of GRB 020819 may be more common than previously anticipated. 

These findings and the clear example of the host galaxy of GRB 020819, show that the hosts of GRBs may be far more diverse, in terms of extinction, mass and metallicity, than previously found.

\begin{acknowledgements}
      We thank Paulo Afonso for his help on carrying out the observations, and the referee for useful comments and suggestions.
      TK acknowledges support by
     the DFG cluster of excellence ``Origin and Structure of the Universe". 
      Part of the funding for GROND (both hardware as well as personnel)
     was generously granted from the Leibniz-Prize to Prof. G. Hasinger
     (DFG grant HA 1850/28-1).  
\end{acknowledgements}

\bibliographystyle{aa}

\end{document}